\documentclass[a4paper,10pt,twoside]{cpc-hepnp}
\usepackage{CJK,upgreek,fancyhdr}
\usepackage{multicol}
\usepackage{graphicx}
\usepackage{booktabs}
\usepackage{amssymb,bm,mathrsfs,bbm,amscd}
\usepackage[tbtags]{amsmath}
\usepackage{lastpage}

\begin{document}
\begin{CJK*}{GB}{gbsn}

\fancyhead[c]{\small Chinese Physics C~~~Vol. xx, No. x (201x) xxxxxx}
\fancyfoot[C]{\small 010201-\thepage}

\title{Warm  Tachyon Inflation and Swampland Criteria}

\author{%
      Abolhassan Mohammadi$^{1,2}$\email{a.mohammadi@uok.ac.ir; abolhassanm@gmail.com}%
\quad Tayeb Golanbari$^{1}$\email{t.golanbari@uok.ac.ir; t.golanbari@@gmail.com}
\quad Haidar Sheikhahmadi$^{3,4}$\email{h.sh.ahmadi@gmail.com;h.sheikhahmadi@ipm.ir}\\
\quad Kosar Sayar$^{1}$\email{sayar1369@yahoo.com}
\quad Lila Akhtari$^{1}$\email{liakhtari@yahoo.com}
\quad M. A. Rasheed$^{2,5}$\email{mariwan.rasheed@uhd.edu.iq}
\quad Khaled Saaidi$^{1}$\email{ksaaidi@uok.ac.ir; khaledsaedi@gmail.com}
}

\maketitle

\address{%
$^1$ Department of Physics, Faculty of Science, University of Kurdistan, Pasdaran St. P.O. Box 66177-15175, Sanandaj, Iran\\
$^2$Development Center for Research and Training (DCRT), University of Human Development, Sulaimani,
Kurdistan Region, Iraq,\\
$^3$Center for Space Research, North-West University, Mafikeng, South Africa,\\
$^4$School of Astronomy, Institute for Research in Fundamental Sciences (IPM), P. O. Box 19395-5531, Tehran, Iran\\
$^5$Department of Physics, College of Science, University of Sulaimani, Sulaimani, Kurdistan Region, Iraq \\
}

\begin{abstract}
The scenario of two components warm tachyon inflation is considered where the tachyon field plays the role of inflaton and drives inflation. During inflation, the tachyon scalar field interacts with the other component of the Universe which is considered as photon gas, i.e. radiation.
The interacting term contains a dissipation coefficient, and the study is modeled based on two different and familiar choices of the coefficient that have been \textbf{studied} in the literature. By applying the latest observational data,  the acceptable ranges for the free parameters of the model are obtained. For any choice inside the \textbf{estimated} ranges, there is an acceptable concordance between the theoretical predictions and observations. Whereas the model is established based on some assumptions, it is vital to check their validity for the obtained values of the free parameters of the model. It is realized that the model is not self-consistent for all values of the ranges and sometimes the assumptions are violated. Therefore, \textbf{to have both self-consistency and agreement with data} the parameters of the model need to be \textbf{constrained} again. After that, we are going to consider the recently proposed swampland conjecture, which imposes two conditions on the inflationary models. These criteria could rule out some of the inflationary models, however, warm inflation is known as one of those models that could successfully satisfy the swampland criteria. A precise investigation determines that the proposed warm tachyon inflation could not satisfy the swampland criteria for some cases. In fact, for the first case of the dissipation coefficient, where there is dependency only on the scalar field, the model could agree with observational data, however, it is in direct tension with the swampland criteria. But, for the second case where the dissipation coefficient has a dependency on both scalar field and temperature, the model shows an acceptable agreement with observational data and it could properly satisfy the swampland criteria.

\end{abstract}




\begin{multicols}{2}

\section{Introduction}
Inflationary scenario is famed as one of the best proposals of describing the universe's evolution at very early eras. The first proposal of inflation was introduced by A. Starobinsky based on a conformal anomaly in quantum gravity \cite{starobinsky1980new}. The main goal of the model was to solve the question of the initial singularity, and it was built based on the assumption of the stage of a quasi-de Sitter in the very early Universe \cite{Linde:2000kn,Linde:2005ht,Linde:2005vy,Linde:2004kg}. This model showed a graceful exit from the inflationary stage, and in this regard, it could be counted as the first model of inflation. The model played an important role in developing the scenario of inflation \cite{Linde:2000kn,Linde:2005ht,Linde:2005vy,Linde:2004kg}. One year later, an inflationary model was introduced by A. Guth aiming to solve the problems of the hot big-bang theory \cite{Guth:1980zm}. This scenario, which is known as old inflation, was suffered from the bubble nucleation problem. However, the idea was very simple and elegant which had a deep impact on the future cosmological inflationary models. The new inflationary scenario \cite{albrecht1982cosmology,linde1982new} could properly solve the problem of Guth's model, where the scalar field stands at the top of the effective potential and then it slowly rolls down to the bottom. \textbf{In contrast} to the old inflation where inflation occurs at the false vacuum with $\dot{\phi}=0$, here the stage of inflation happens during a slowly rolling of the inflaton toward the minimum of its potential, i.e. $\dot{\phi} \neq 0$ \cite{Linde:2000kn,Linde:2005ht,Linde:2005vy,Linde:2004kg}. The main problem of the new inflation is that the density perturbations that are generated during inflation are very large and consequently unacceptable. This problem is avoided by using a small coupling constant of the scalar field. However, for small coupling constant the scalar field could no longer be in \textbf{the state} of the thermal equilibrium with other matter fields \cite{Linde:2000kn,Linde:2005ht,Linde:2005vy,Linde:2004kg}. Complete modification of the big-bang theory was presented by \textbf{the invention} of the scenario of Chaotic inflation \cite{linde1983chaotic}, which could solve the problems of both old and new inflations. An interesting feature of the scenario is that \textbf{inflation} could happen even for a simple potential like $V \propto \phi^n$. \\
After that, many inflationary scenarios have been proposed in which non-canonical inflation \cite{Barenboim:2007ii,Franche:2010yj,Unnikrishnan:2012zu,Gwyn:2012ey,Rezazadeh:2014fwa,Cespedes:2015jga,Stein:2016jja,
Pinhero:2017lni,Teimoori:2017wbx,Mohammadi:2019dpu}, tachyon inflation \cite{Fairbairn:2002yp,Mukohyama:2002cn,Feinstein:2002aj,Padmanabhan:2002cp}, DBI inflation \cite{Spalinski:2007dv,Bessada:2009pe,Weller:2011ey,Nazavari:2016yaa,Amani:2018ueu,Mohammadi:2018zkf} G-inflation \cite{maeda2013stability,abolhasani2014primordial,alexander2015dynamics,tirandari2017anisotropic}, brane inflation \cite{maartens2000chaotic,golanbari2014brane} could be addressed as some of them. \textbf{All of these} scenarios have similar assumptions. The scalar field is the dominant component at the time, and inflation happens during a slowly rolling of the scalar field from the top toward the minimum of the potential. It is worth mentioning that applying the idea of inflation in the Starobinsky model leads to \textbf{great} achievement in which the final result has a great consistency with observational data. The model is known as $R^2$  Starobinsky-inflation \cite{starobinsky1980new}. The process of particle creation and heating up the universe happen at the end of inflation during preheating and reheating stages, where the scalar field oscillates around the minimum of its potential with time scales shorter than the Hubble time and its energy is drained to other matter fields, for instance, radiation \cite{Kofman:1994rk}. \\
In 1995, a new picture of inflation was introduced by A. Berera which is known as warm inflation \cite{berera1995warm}. According to the warm inflation, the scalar field is still the dominant component of the universe,\textbf{ however,} the interaction between \textbf{the scalar} field and other fields is not ignored. Due to the interaction, the energy is transferred from the scalar field to the radiation. Therefore, there is a particle production mechanism during inflation, and the Universe's temperature \textbf{does} not suddenly drop. The Universe remains warm and full of other particles in which the scenario of reheating is no longer required, and the universe smoothly enters to the radiation era \cite{berera1995warm,berera2000warm,taylor2000perturbation,Dymnikova:2001ga,Dymnikova:2001jy,hall2004scalar,BasteroGil:2004tg,Sayar:2017pam,Akhtari:2017mxc,Sheikhahmadi:2019gzs,Harko:2020cev}. Another difference stands in the type of the cosmological perturbations. In warm inflation there are both quantum and thermal fluctuations, however, thermal fluctuations are the dominant ones. The thermal fluctuations are proportional to the fluid temperature $T$ and the quantum fluctuations are proportional to the Hubble parameter $H$. Then, the condition of domination of thermal fluctuation leads to the $T>H$ inequality condition. \\
There are many \textbf{works of literature} which have devoted to \textbf{studying} different aspects of warm inflationary scenario for different models \cite{berera1995warm,berera2000warm,taylor2000perturbation,hall2004scalar,BasteroGil:2004tg,Sayar:2017pam,Akhtari:2017mxc}. In the present work, we are going to reconsider the warm inflation where the tachyon field plays the role of inflaton. The main motivations for the present work are stated in the following lines. The first reason is the \textbf{recently} proposed swampland conjecture. The recent studies on the effective field theory (EFT) and string theory lead to two swampland criteria \cite{Obied:2018sgi,Ooguri:2018wrx}: I) imposing an upper bound on the field range, i.e. $\Delta\phi < \Delta$ (where $\Delta$ is of \textbf{the order} of unity), which rises from this belief that the effective Lagrangian in the EFT is valid only for a finite radius; II) putting an upper bound on the gradient of the potential of the field of any EFT, i.e. $|V'|/V \geq c$ (where the most recent studies \textbf{determine} that $c$ could be even of \textbf{the order} of $\mathcal{O}(0.1$) \cite{Kehagias:2018uem}). The second criterion implies that the first slow-roll parameter, i.e. $\epsilon_{\phi} = (V' / V)^2,$ and the tensor-to-scalar ratio, $r = 16\epsilon_{\phi} > 8c^2,$  should not be small. The desire for satisfying these two criteria could rule out some of the inflationary models, however, there \textbf{is still a possibility} for some other models to survive \cite{Kehagias:2018uem,Das:2018hqy,Das:2018rpg,Goswami:2019ehb,Kinney:2018kew,Matsui:2018bsy,Lin:2018rnx,Dimopoulos:2018upl,
Kinney:2018nny}. The k-essence model \cite{ArmendarizPicon:1999rj,Garriga:1999vw,Mohammadi:2019qeu} is one of them, where $r=16 c_s \epsilon_{\phi}$ and the sound speed $c_s$ could be smaller than unity \cite{Garriga:1999vw}. Tachyon model, which has been inspired from string theory, is known as a subclass of k-essence model, and could be a suitable choice for considering the swampland criteria. The other inflationary model which is able to survive the aforementioned criteria is warm inflationary scenario where the first slow-roll parameter is obtained as $\epsilon = \epsilon_{\phi}/Q$ and tensor-to-scalar ratio is found as $r = (H / T) \; (16 \epsilon_\phi / (1+Q)^{5/2})$ \cite{Berera:1999ws,Berera:2004vm,BasteroGil:2009ec,Bartrum:2013fia,Bastero-Gil:2016qru,Visinelli:2011jy,BasteroGil:2012zr,
Bastero-Gil:2013nja,Visinelli:2014qla,Bastero-Gil:2018uep}. The parameter $Q$ is known as the dissipative parameter which in strong dissipative regime is bigger than unity, $Q \gg 1$. This feature helps the scenario to successfully pass the criteria and satisfy them \cite{Das:2018hqy,Das:2018rpg,Goswami:2019ehb,Motaharfar:2018zyb,Das:2019hto,Kamali:2019xnt,Berera:2019zdd,DallAgata:2019yrr,
Brandenberger:2020oav}. The second motivation is related to the importance of the tachyon field. After the introducing of tachyon model in cosmological studies \cite{Sen:2002nu,Sen:2002in,Sen:2002an,Gibbons:2002md}, the model has received a huge attention and found a place in all area of cosmology including the inflation \cite{Fairbairn:2002yp,Mukohyama:2002cn,Feinstein:2002aj,Padmanabhan:2002cp,Aghamohammadi:2014aca,Mohammadi:2018oku}. \\
The warm inflation including the tachyon field as inflaton has been studied in \cite{Herrera:2006ck,delCampo:2008fc,Deshamukhya:2009wc,Zhang:2013waa,Cid:2015ota,Bilic:2013jk,Setare:2014uja,Kamali:2015yya,Kamali:2016frd,Motaharfar:2016dqt}, however we are going to reconsider the scenario with a different interaction terms and also working with the Hubble parameter instead of the potential namely Hamilton-Jacobi formalism \cite{Salopek:1992qy,Liddle:1994dx,Kinney:1997ne,Guo:2003zf,Aghamohammadi:2014aca,Saaidi:2015kaa,Sheikhahmadi:2016wyz}.
The study of the inflationary models are usually performed using three methods:
  \begin{enumerate}
    \item introducing the potential: which is the most common method in the inflationary studies
    \item Introducing the Hubble parameter: here, instead of the potential, the Hubble parameter is introduced as a function of the scalar field. The method is known as the Hamilton-Jacobi formalism
    \item Introducing the scale factor: in this method the scale factor is introduced as a function of time, \emph{e.g. }the intermediate inflation
  \end{enumerate}
The first approach imposes some restrictions on the form of the potential and evolution of the scalar field. However, in the second approach, there are some conditions on the evolution of the Hubble parameter. In general, the Hamilton-Jacobi formalism provides a clear geometrical interpretation and more convenient analysis. Some features of the formalism could be addresses as: 1) More accurate expressions for the slow-roll parameters, 2)Neglecting extra assumptions, 3) easy to work with (detail explanation about the formalism and its features could be found in \cite{Liddle:1994dx} and references therein).  \\
The main focuses of the present work are on considering the consistency of the model with observational data and qualitatively considering its agreement with the swampland criteria as well. Following \cite{BasteroGil:2011xd}, the interaction term in the conservation equations is assumed to include a dissipation coefficient and a sum of energy density and pressure of the scalar field. In this regard, the interaction is different with the other performed works on the topic; there is $H^2 \Gamma \dot\phi^2$ instead of the usual term $\Gamma \dot\phi^2$. Another consequence of the selected interaction term is that we have the same definition for the dissipative parameter $Q=\Gamma/3H$, independent of the type of the scalar field.

The dissipation coefficient can be considered as a function of either the scalar field or temperature, and in some cases, it depends on both \textbf{the scalar field} and the temperature. Here, two different choices are picked out for this coefficient. At first, it is assumed that there is only \textbf{dependency on} the tachyon field, and in the second case, it will be considered as a function \textbf{of both} the tachyon field and fluid temperature. The main perturbation parameters are obtained for the model, and in comparison with the latest observational data, it is tried to determine the free parameters of the model. In the next step, the self-consistency of the model is considered. The model is constructed based on some assumptions, and we are going to examine their validity for the obtained values of the constants of the model. The self-consistency of the model is an important point that is missing in lots of works. \\
The paper is organized as follows: In Sec.2, the dynamical perturbation equations of the model are discussed. Then, \textbf{in} the last part of the section, we are going to rewrite the equations for the strong dissipative regime. To compare the model with observational data, two different choices for the dissipation coefficient will \textbf{be picked} out in Sec.3, and they will be investigated in separately. For each case, the free constants of the model are specified using the observational data, and for each case, we consider \textbf{the consistency} of the results with the main conditions of the model. In Sec.4, the swampland criteria for the model \textbf{are} discussed. The results of the model are summarized in the conclusion section.

\section{Tachyon Model}\label{model}
It is assumed that the Universe is filled with a scalar field, that drives inflation (named inflaton), and photon gas. The geometry of the Universe is described by a spatially flat FLRW metric. Then, the Friedmann equation is given as
\begin{equation}\label{friedmann}
  3H^2 = \rho_\phi + \rho_r
\end{equation}
where $M_p^2=1/8\pi G = 1$. As mentioned in the introduction, the inflton is taken to be a tachyon fluid which can be described by a diagonal energy-momentum-tensor, $T_{\phi \nu}^{\mu}=diag(-\rho_\phi,p_\phi,p_\phi,p_\phi)$\footnote{here, the signature of the metric is $+2$.} \cite{Gibbons:2002md}. It is easy to show that the energy density and pressure are given by following relations respectively \cite{Gibbons:2002md}
\begin{equation}\label{tachyonrhop}
\rho_\phi=\dfrac{V(\phi)}{\sqrt{1-\dot{\phi}^2}}, \; \qquad \; p_\phi=-V(\phi)\sqrt{1-\dot{\phi}^2},
\end{equation}
where dot denotes derivative with respect to the cosmic time $t$, and $V(\phi)$ refers the potential of the tachyon field. \\
Due to the interaction between the tachyon field and radiation, the energy conservation equations are modified as \cite{BasteroGil:2011xd}
\begin{eqnarray}
  \dot{\rho}_{\phi} + 3H(\rho_\phi+p_\phi) & = &- \Gamma \; (\rho_\phi+p_\phi), \label{phiconservation} \\
  \dot{\rho}_{r} + 3H(\rho_r+p_r) & = & \Gamma \; (\rho_\phi+p_\phi). \label{rconservation}
\end{eqnarray}
in which $\Gamma$ is the dissipation coefficient in which in its general form it can be a function of the both of scalar field and temperature. The radiation part has a well-known equation of state as $p_r = \rho_r /3$.
Using Eq.\eqref{tachyonrhop} and Friedmann equation \eqref{friedmann}, the interaction term is obtained as
\begin{equation*}
  \Gamma \; (\rho_\phi+p_\phi) = 3\Gamma H^2 \dot{\phi}^2
\end{equation*}
which is different from those interaction terms that have been introduced in \cite{Herrera:2006ck,delCampo:2008fc,Zhang:2013waa,Cid:2015ota,Kamali:2016frd}. This difference leads to the usual definition of the dissipative parameter as $Q \equiv \Gamma / 3 H$. \\
The tachyon field equation of motion is obtained by substituting Eq.\eqref{tachyonrhop} into Eq.\eqref{phiconservation}
\begin{equation}\label{tachyonEoM}
{\ddot{\phi} \over 1-\dot{\phi}^2 } + 3 H \dot{\phi} + {V' \over V} = - \Gamma \dot{\phi},
\end{equation}
where prime denotes derivative with respect to the tachyon field $\phi$. \\

To have an accelerated expansion phase, it is assumed that the tachyon field dominates the photon gas energy density. Then, the Friedmann Equation (\ref{friedmann}) is rewritten as
\begin{equation}\label{friedmann11}
H^2 = \rho_\phi = \;\dfrac{V(\phi)}{\sqrt{1-\dot{\phi}^2}}\;,
\end{equation}
The second Friedmann equation is obtained by taking the time derivative of Eq.\eqref{friedmann11} and using Eq.\eqref{phiconservation}
\begin{equation}\label{thubble}
\dot{H}=\dfrac{-3H^2}{2}\;(1+Q)\;\dot{\phi}^2.
\end{equation}
Assuming that the Hubble parameter is a function of the tachyon field, i.e. $H:= H(\phi)$, and by using the fact that $\dot{H}:=H'\dot{\phi}$, the time derivative of the field is found as
\begin{equation}\label{phidot}
\dot{\phi}=\dfrac{-2}{3(1+Q)}\;\dfrac{H'}{H^2}\;.
\end{equation}
Another assumption in the scenario of warm inflation is quasi-stable production of the photon gas, i.e. $ \dot{\rho_\gamma}\ll 4H\rho_\gamma , \Gamma\dot{\phi}^2 $, in which by imposing this condition on Eq.\eqref{rconservation}, the radiation energy density is obtained as
\begin{equation}\label{radiattionrho}
\rho_\gamma={3 \over 4} \; H \Gamma = \alpha T^4\;,
\end{equation}
where $T$ is the temperature of thermal bath and $\alpha$ is well-known Stephen-Boltzman constant. \\
The first slow-roll parameter is defined as $\epsilon \equiv - \dot{H} / H^2$ that by using Eq.(\ref{phidot}) it comes to be
\begin{equation}\label{epsilon}
\epsilon(\phi)\equiv -\dfrac{\dot{H}}{H^2}=\dfrac{2}{3(1+Q)}\;\dfrac{H'^2}{H^4}\;,
\end{equation}
The second slow-roll parameter is $\epsilon_2\equiv-{\dot{\epsilon}}/{H \epsilon}$ which after some manipulations, one arrives at
\begin{equation}
\epsilon_2=\left( 2\epsilon(\phi) - 2\eta(\phi) \right) + {Q \over 1+Q}\left(\epsilon(\phi) - \beta(\phi)\right)\;,
\end{equation}
so that the parameter $\eta$ and $\beta$ are expressed as follows
\begin{equation*}
\eta(\phi)\equiv \dfrac{4}{3(1+Q)}\;\dfrac{H''}{H^3}, \qquad \beta(\phi)\equiv \dfrac{2}{3(1+Q)}\;\dfrac{\Gamma' H'}{\Gamma H^3}.
\end{equation*}
Whereas Hubble parameter is given in terms of the tachyon scalar field, one can extract the potential of the model of Friedmann equation (\ref{friedmann11})  viz.,
\begin{equation}\label{potential}
V(\phi)=3M_P^2 H^2(\phi)\sqrt{1-\dfrac{2}{3}\;\dfrac{1}{(1+Q)}\epsilon(\phi)}\;,
\end{equation}
where to receive this expression the Eqs.\eqref{phidot} and \eqref{epsilon} have been applied. \\
The amount of the Universe expansion during the inflationary times is measured by the number of e-folds, $N$, defined as\footnote{The right hand side should be written as $\int dN = Ne - N_{\star}$ in which the subscripts "$e$" and "$\star$" respectively stand for end of inflation and horizon crossing time. To solve the horizon and flatness problems it is belived that there should be about $60-65$ number of e-fold. Then, for the rest of the paper, the $N_\star$ is put to $N_\star = 0$ and $N_e = 65$.}
\begin{equation}\label{N}
N=\int\dfrac{H}{\dot{\phi}}d\phi=\dfrac{3}{2}\int_{\phi_e}^{\phi}(1+Q)\;\dfrac{H^3}{H'}d\phi\;,
\end{equation}
where the subscript "e" in $\phi_e$ denotes the tachyon field at the end of inflation. \\
Besides the evolution of the background parameters, we need to know about the perturbative behaviours of the parameters at the inflationary period. Cosmological perturbations are known as one of the most important predictions of the inflation. These perturbations can in general be divided into three types as scalar, vector and tensor ones. The vector type is usually ignored due to the fact that it depends on the inverse of the scale factor and will be diluted exponentially during inflation. The primordial seeds of large scale structure of the Universe is believed to be the scalar perturbations that are produced during inflation. In the warm inflationary scenario, there are both quantum and thermal fluctuations, but, it is assumed that the thermal fluctuations \textbf{overcome} the quantum fluctuations. The power-spectrum of these fluctuations for the tachyon model \textbf{is} given by \cite{Herrera:2006ck}
\begin{equation}\label{ps}
\mathcal{P}_s= { exp(-2\chi(\phi)) \over \big( V' /V \big)^2 } \; \delta\phi^2\;,
\end{equation}
in which $\delta \phi$ is the fluctuation in the scalar field, and the parameter $\chi$ is defined as
\begin{eqnarray}\label{chi}
\chi(\phi) & = & \int \left[\dfrac{1}{(3H+\tilde{\Gamma}/V)}\;\left(\dfrac{\tilde{\Gamma}}{V} \right)'
+\dfrac{9}{8}\;\dfrac{(\tilde{\Gamma}/V+2H)}{(\tilde{\Gamma}/V+3H)^2}\times \right. \qquad \nonumber \\
  & &  \qquad \qquad \left(\tilde{\Gamma}+4 HV-\dfrac{\tilde{\Gamma}' (V'/V)}{12H(3H+\tilde{\Gamma}/V)} \right) \times \nonumber \\
   &  & \qquad \qquad \qquad \qquad  \dfrac{V'/V}{V} \; \Bigg] d\phi\;.
\end{eqnarray}
For our model $\tilde{\Gamma}$ is equal to $3 \Gamma H^2$, resulted from Eq.\eqref{tachyonEoM}. The scalar spectral index, defined as $\mathcal{P}_s = \mathcal{P}_s^\star \big( k / k_\star \big)^{n_s-1}$, is another observational parameter that measures the scale-dependency of the power-spectrum, in which $n_s=1$ indicates the power-spectrum is scale invariant. This parameter is obtained by taking log-derivative of the power-spectrum as
\begin{equation}\label{ns}
n_s-1 = {d\ln\big( \mathcal{P}_s \big) \over d\ln k}.
\end{equation}
Another type of primordial fluctuations are the tensor one, which also known as the primordial gravitational waves. Since the fluid has no role in the tensor perturbations equation, the power-spectrum of tensor perturbations is obtained as same as the cold inflationary scenario, i.e. $\mathcal{P}_t= H^2 / 2\pi^2$ \cite{delCampo:2008fc,Zhang:2013waa,Cid:2015ota,Zhang:2009ge,del2007cosmological}. The tensor perturbations are measured indirectly through the parameter $r$ which is defined as the ratio of the power-spectrums of the tensor perturbation to the scalar perturbations, $ r = \mathcal{P}_t / \mathcal{P}_s$. In contrast to the scalar perturbations, there is no exact data for this parameter and there is only an upper bound $r > 0.064$ .


\subsection{Strong Dissipative Regime}
Depend on the value of the dissipative parameter $Q$, the study of warm inflation could be divided into two regimes as strong dissipative regime (SDR) and weak dissipative regime (WDR); respectively correspond to $Q \gg 1$ and $Q \ll 1$. For the rest of the work, the model is considered only for SDR where one can use the approximation $(1+Q) \simeq Q$ in the equations. \\
In warm inflation, thermal fluctuations dominate over quantum fluctuations and the corresponding fluctuations in scalar field is given by $\delta\phi^2\approx {k_F T}/{2\pi^2}$ where $k_F=\sqrt {\tilde{\Gamma} H/V}$ \cite{Herrera:2006ck}. Plugging this into Eq.(\ref{ps}), the amplitude of the scalar perturbations in SDR becomes
\begin{equation}\label{pssdr}
\mathcal{P}_s= { exp(-2\tilde{\chi}(\phi)) \over \big( 2H'/H \big)^2 } \; {T \over 2\pi^2} \; \sqrt{H},
\end{equation}
and the defined parameter $\chi$ is reduced to
\begin{eqnarray}\label{chisdr}
\tilde{\chi}(\phi) & = & \int \left[ {\Gamma' \over \Gamma} + {9 \over 8} \; {1 \over 3HQ}
\left( 3H^2 Q - {(3H^2 \Gamma)' (2H'/H) \over 36 H^2 Q} \right) \times \right. \nonumber \\
    &  & \left. \qquad \qquad \qquad \qquad \; {(2H'/H) \over 3H^2} \right] d\phi\;.
\end{eqnarray}
The scalar spectral index $n_s$, that is related to the power-spectrum of the scalar perturbation via \eqref{ns}, is obtained in terms of the slow-roll parameters as
\begin{equation}\label{nssdr}
n_s - 1 = {13 \over 4} \; \epsilon(\phi) + {3 \over 2} \; \eta(\phi) + {7 \over 4} \; \beta(\phi).
\end{equation}
One notices  that the spectral index is obtained up to the first order of slow-roll parameters. \\
Using Eq.\eqref{pssdr} and tensor power-spectrum, the tensor-to-scalar ratio is obtained as
\begin{equation}\label{rsdr}
r = {16 H'^2 \over T \sqrt{H}} \; exp(2\tilde{\chi}(\phi))
\end{equation}
In the next section, two typical examples are considered for the dissipative coefficient $\Gamma$. Also, the Hubble parameter is assumed as a power-law function of scalar field, i.e. $H(\phi)=H_0 \phi^n$ for the rest of the work. \\

\section{Consistency with Observation}\label{data}
To check the accuracy and consistency  of any inflationary models it is required to compare  theoretical predictions with observation, e.g. \cite{Planck:2013jfk,Ade:2015lrj,Akrami:2019izv}. In this regard, the dissipative coefficient $\Gamma$ should be specified which can be considered as a function of the scalar field, or in more general case it could be a function of both the scalar field and fluid temperature. In the following subsections, we are about to consider both cases. \\

\subsection{First Case: $\Gamma = \Gamma_0 \phi^m$}
As the first case, the dissipation coefficient is taken as a power-law function of tachyon field, i.e. $\Gamma = \Gamma_0 \phi^m$, in which $\Gamma_0$ and $m$ are constants. From Eq.(\ref{phidot}), the time derivative of the tachyon field becomes
\begin{equation}\label{phidotsdr01}
\dot{\phi}=-\dfrac{2 n}{\Gamma_0}\;\dfrac{1}{\phi^{m+1}}\;.
\end{equation}
The first slow-roll parameter $\epsilon$  is determined completely by substituting the introduced dissipation function into \eqref{epsilon}. Obviously inflation ends as slow-roll parameter $\epsilon$ reaches unity, therefore the scalar field is read as
\begin{equation}\label{phie01}
\phi_e^{m+n+2} = {2 n^2 \over H_0 \Gamma_0} \; .
\end{equation}
The scalar field at the time of horizon exit is obtained through the number of e-fold \eqref{N} as
\begin{equation}\label{phis01}
\phi_\ast^{m+n+2}=\phi_e^{m+n+2} \; \left( 1 + {(m+n+2) \over n} \; N \right)\;.
\end{equation}
where $\star$ indicates the time of horizon crossing. The slow-roll parameters, at this edge, are obtained as
\begin{eqnarray}
\epsilon^\ast & = & \left( 1 + {(m+n+2) \over n} \; N \right)^{-1} \equiv \bar{N}^{-1} \;, \label{srpstare} \\
\eta^\ast & = & {(n-1) \over n} \; \epsilon^\ast\; ,\label{srpstaret} \\
\beta^\ast & = & {m \over n} \; \epsilon^\ast\; . \label{srpstarb}
\end{eqnarray}
Applying these results to Eq.\eqref{nssdr}, the scalar spectral index at the time of horizon exit is given by
\begin{equation}\label{nssdr01}
  n_s - 1 = \left( {13 \over 4}  + {3 \over 2} \; {n-1 \over n} + {7 \over 4} \; {m \over n} \right) \; \bar{N}^{-1}\;,
\end{equation}
one should note that the scalar spectral index only depends on the constants $n$ and $m$. On the other hand, by integrating \eqref{chisdr} and substituting the result into Eq.\eqref{pssdr}, the power-spectrum reads
\begin{equation}\label{pssdr01}
  \mathcal{P}_s = {1 \over 8\pi^2 n^{3/2}} \; \left( {\Gamma_0 \over 3\alpha H_0} \right)^{1/4} \;
   {\exp\left[ - {3(2n+m) \over 8 (m+n+2)} \; \epsilon(\phi) \right] \over \phi^{7m+19n-6 \over 4}},
\end{equation}
The tensor-to-scalar ratio is easily derived, so that
\begin{equation}\label{rsdr01}
  r = 4 n^{3/2} \; \left( {3\alpha H_0^9 \over \Gamma_0} \right)^{1/4} \; \phi^{7m+27n-6 \over 4} \; \exp\left[ {3(2n+m) \over 8 (m+n+2)} \; \epsilon(\phi) \right]
\end{equation}
Then by using Eqs.\eqref{phis01} and \eqref{srpstare}, the power spectrum and tensor-to-scalar ratio are obtained at the time of horizon crossing. \\
The first conclusion that could be made from Eqs.\eqref{pssdr01} and \eqref{rsdr01} is that the constant $n$ must be positive, otherwise there will be an imaginary value for the parameters $\mathcal{P}_s$ and $r$, which is unphysical. \\
Now, to constrain the free parameters of the model, the results at the time of horizon crossing should be examined with data. At this time, the scalar spectral index depends on both parameters $n$ and $m$. The situation is different for tensor-to-scalar ratio in which besides $n$ and $m$ the other two constants $\Gamma_0$ and $H_0$ also appear in definition of $r(t=t_\star)$. Based on Planck data, there is an exact value for the amplitude of the scalar perturbations, and there are also some statements about the energy scale of inflation. Thus, From the energy scale of inflation, Eq.\eqref{potential}, the constant $H_0$ is determined  as\footnote{During the inflation, the slow-roll $\epsilon$ is smaller than unity and also the dissipative parameter $Q$ is large because we are standing in SDR. Then, the second term in Eq.\eqref{potential} could be ignored with a good approximation.}
\begin{equation}\label{H0-01}
  H_0 = \bar{V}^\star \; \Gamma_0^{n \over m+2}, \qquad  \bar{V}^\star \equiv { \left( V^\star / 3 \right)^{m+n+2 \over 2(m+2)} \over \left( 2n^2 \bar{N} \right)^{n \over m+2} },
\end{equation}
in which $V^\star$ is the energy scale of inflation. Substituting the obtained $H_0$ in the amplitude of the scalar perturbations, $\Gamma_0$ is extracted as
\begin{equation}\label{G0-01}
  \Gamma_0 = \left( \mathcal{P}_s^\star \over D \right)^{4(m+2) \over 8m+18n-4}
\end{equation}
where the parameter $D$ is defined as
\begin{equation*}
  D = {1 \over 8\pi^2 n^{3/2}} \; {1 \over \big( 3\alpha \bar{V}^\star \big)^{1/4}} \;  { \exp\left[ - {3(2n+m) \over 8 (m+n+2) \; \bar{N}} \;  \right] \over \big( 2n^2 \bar{N} \big)^{7m+19n-6 \over 4(m+n+2)}} \;
\end{equation*}
From Eqs.\eqref{H0-01} and \eqref{G0-01}, it is realized that both perturbation parameters $n_s$ and $r$ now depend only on $n$ and $m$. Utilizing Planck $r-n_s$ diagram, a set of points is obtained for $n$ and $m$ in which for any $(n,m)$ point in the set the result of the model is in good consistency with observation. Fig.\ref{Snm01} illustrates this set of points, in which the dark blue color determines the $(n,m)$ points that our model is going to be in agreement with $68\%$ CL of Planck $r-n_s$ diagram, and the light blue color is related to $95\%$ CL. \\
\begin{center}
\includegraphics[width=8cm]{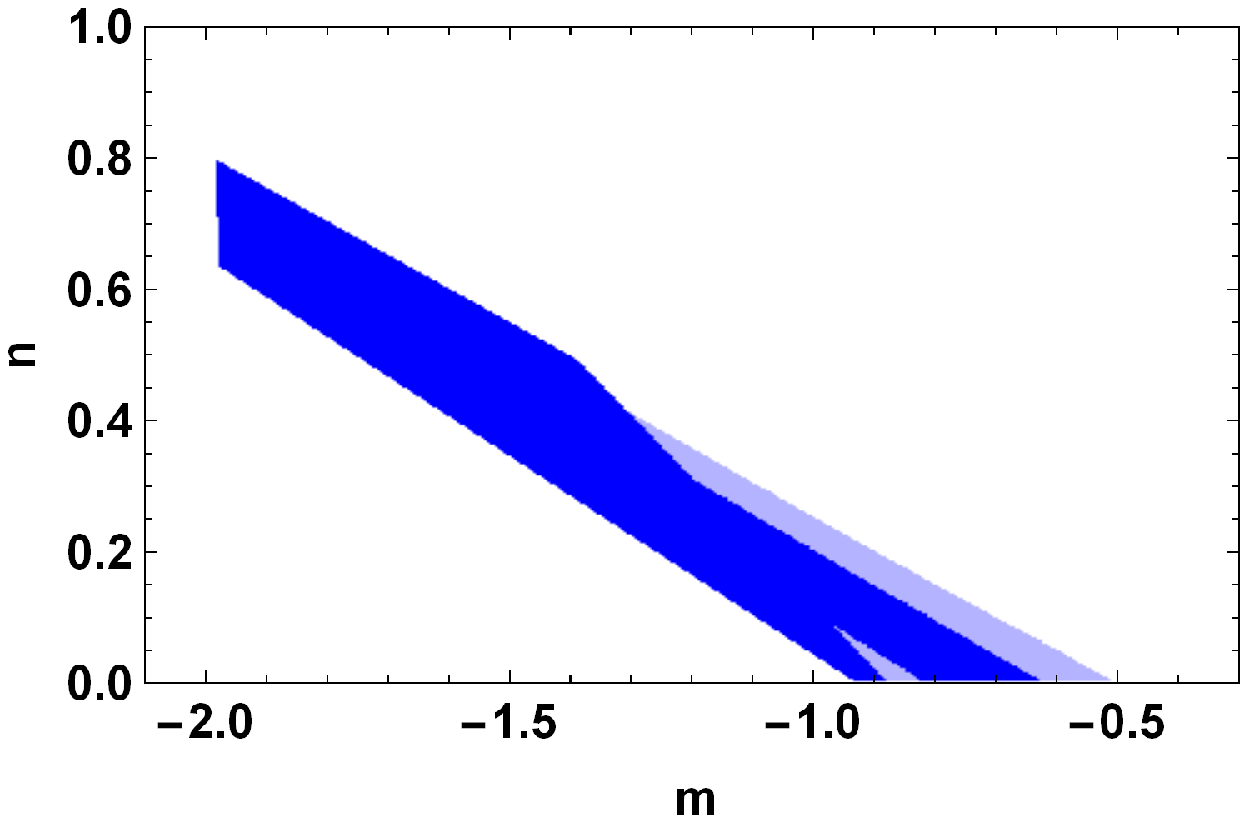}
\figcaption{\label{Snm01} The parametric plot of $(n,m)$.}
\end{center}
It is not the whole story. To build the  model, we made two main assumptions and we are going to find out if for all $(n,m)$ points of Fig.\ref{Snm01} the assumptions are still valid or not. Then, it is very important that these \textbf{assumptions} be verified for the whole time of inflation. In the first postulation, that is in the warm inflationary scenario the thermal fluctuations have to dominate over the quantum fluctuations, described by the condition $T > H$. The second assumption is that the model is restricted to the SDR, where the dissipative parameter is larger than unity, i.e. $Q > 1$. Therefore, we are interested only in the values of $(n,m)$ that satisfactorily pass the conditions and simultaneously  put the model in agreement with data. These values have been depicted in Fig.\ref{SnmQTH01}.
\begin{center}
\includegraphics[width=8cm]{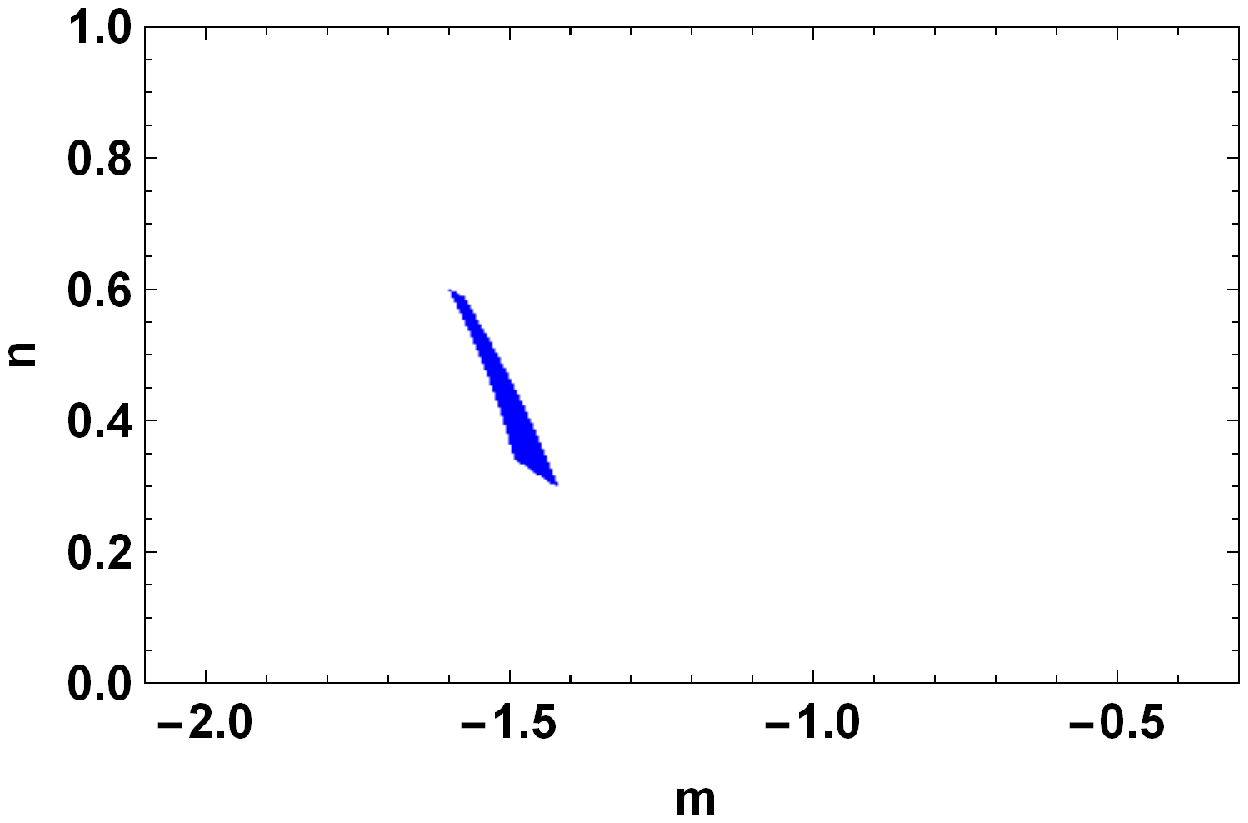}
\figcaption{\label{SnmQTH01} The parametric plot of $(n,m)$ in which for each point of this area the model perfectly meet the observational data. Besides one can show that the main assumptions of the model will be satisfied properly.}
\end{center}
It is clearly seen that even though at the first step we could find a bigger range for the parameter $n$ and $m$, however, the range is tightened by imposing the model conditions. The final result shows that only for a small range of the parameter $(n,m)$ the model comes to an agreement with data and at the same time satisfies the aforementioned conditions. \\
The behavior of $T/H$ and the dissipative parameter $Q$ are depicted in Figs.\ref{THQ01a} and \ref{THQ01b} for different values of $n$ and $m$. It is realized that by passing time and approaching to the end of inflation both of these parameters show increasing in amount. \\
\begin{center}
\includegraphics[width=8cm]{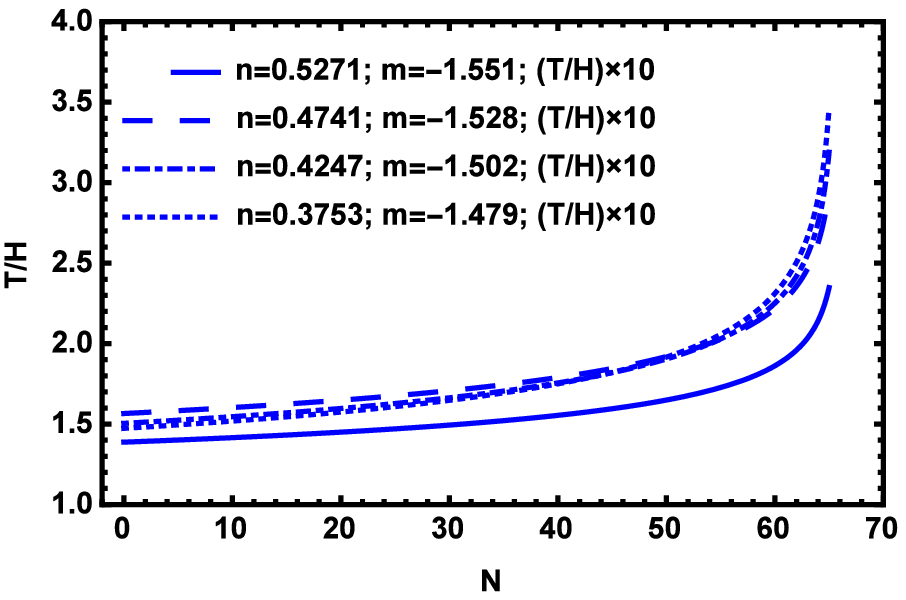}
\figcaption{\label{THQ01a} Behavior of $T/H$ during inflation for different values of $n$ and $m$ selected from Fig.\ref{SnmQTH01}. The plots indicate that both conditions $T/H > 1$ and $Q \gg 1$ are perfectly satisfied.}
\end{center}

\begin{center}
\includegraphics[width=8cm]{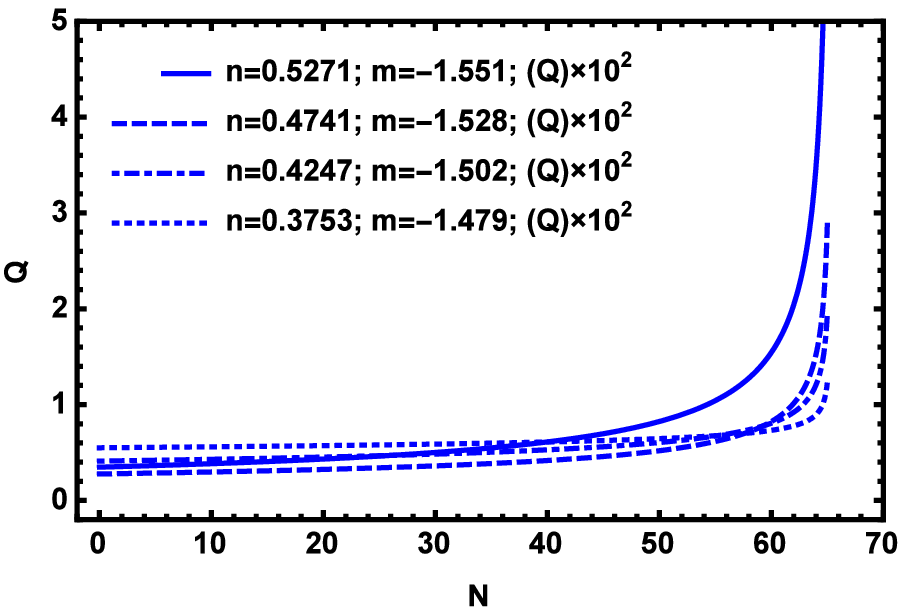}
\figcaption{\label{THQ01b} Behavior of $Q$ during inflation for different values of $n$ and $m$ selected from Fig.\ref{SnmQTH01}. The plots indicate that both conditions $T/H > 1$ and $Q \gg 1$ are perfectly satisfied.}
\end{center}
Table.\ref{Ts01} gives the numerical results for the main perturbation parameters, $T/H$, and also dissipative parameter $Q$ for different values of $n$ and $m$, presented in Fig.\ref{SnmQTH01}. \\
\end{multicols}

\begin{center}
\tabcaption{\label{Ts01} numerical results of the case.}
\footnotesize
\begin{tabular}{cccccccccc}
\hline
    $n$ & $m$ & $\phi_\star$ & $\phi_e$ & $\Gamma_0$ & $H_0$ & $n_s$ & $r$ & $T/H$ & $Q$ \\
    \hline
    $0.5765$ & $-1.582$ & $105.91$ & $0.9120$ & $1.06 \times 10^{3}$ & $6.86 \times 10^{-4}$ & $0.9765$ & $1.27 \times 10^{-8}$ & $14.51$ & $21.91$ \\
    $0.5271$ & $-1.551$ & $89.43$ & $0.6551$ & $9.96\times 10^{2}$ & $8.42 \times 10^{-4}$ & $0.9732$ & $5.43 \times 10^{-9}$ & $13.87$ & $34.70$ \\
    $0.4741$ & $-1.528$ & $105.14$ & $0.6093$ & $8.14 \times 10^{2}$ & $8.81 \times 10^{-4}$ & $0.9689$ & $1.10 \times 10^{-9}$ & $15.65$ & $27.61$ \\
    $0.4247$ & $-1.502$ & $90.33$ & $0.4193$ & $7.62 \times 10^{2}$ & $1.05 \times 10^{-3}$ & $0.9650$ & $4.53 \times 10^{-10}$ & $15.04$ & $41.05$ \\
    $0.3753$ & $-1.479$ & $81.04$ & $0.2891$ & $6.99 \times 10^{2}$ & $1.22 \times 10^{-3}$ & $0.9607$ & $1.70 \times 10^{-10}$ & $14.73$ & $54.94$ \\
    $0.3295$ & $-1.448$ & $52.73$ & $0.1506$ & $7.46 \times 10^{2}$ & $1.54 \times 10^{-3}$ & $0.9572$ & $1.64 \times 10^{-10}$ & $12.22$ & $140.2$ \\
    \hline
    $1$ & $2$ & $52.07$ & $16.36$ & $1.60 \times 10^{-2}$ & $1.06 \times 10^{-4}$ & $1.0207$ & $4.23 \times 10^{7}$ & $10.00$ & $2631$ \\

    $2$ & $1$ & $60.90$ & $21.97$ & $4.77 \times 10^{-1}$ & $3.26 \times 10^{-6}$ & $1.0298$ & $6.75 \times 10^{13}$ & $9.15$ & $799.4$ \\

    $2$ & $0$ & $381.86$ & $112.87$ & $4.37 \times 10^{-1}$ & $1.12 \times 10^{-7}$ & $1.0305$ & $9.82 \times 10^{16}$ & $20.82$ & $8.88$ \\
    \hline
  \end{tabular}

\end{center}

\begin{multicols}{2}
The last three rows of the table are related to the choices of $n$ and $m$ that we have in the canonical cases. As it could be realized from Fig.\ref{SnmQTH01}, these values are out of the range, thence the results are not in consistency  with observational data. The Table.\ref{Ts01} represents the numerical result where one could find that the scalar spectral index is larger than unity, and \textbf{the tensor-to-scalar} ratio is very large, and confirms our first conclusion. \\

It is crucially important for any inflationary model to check whether inflation ends at all. In this regard, the evolution of the slow-roll parameter $\epsilon$ is considered. Fig.\ref{epsilonC01} portrays the behavior of $\epsilon$ versus the number of e-fold for different values of $n$ and $m$. The plot states that $\epsilon$ grows up by passing time and approaching to the end of inflation. Eventually it arrives at one stating that inflation ends and the universe exits from the accelerated expansion phase.
\begin{center}
\includegraphics[width=8cm]{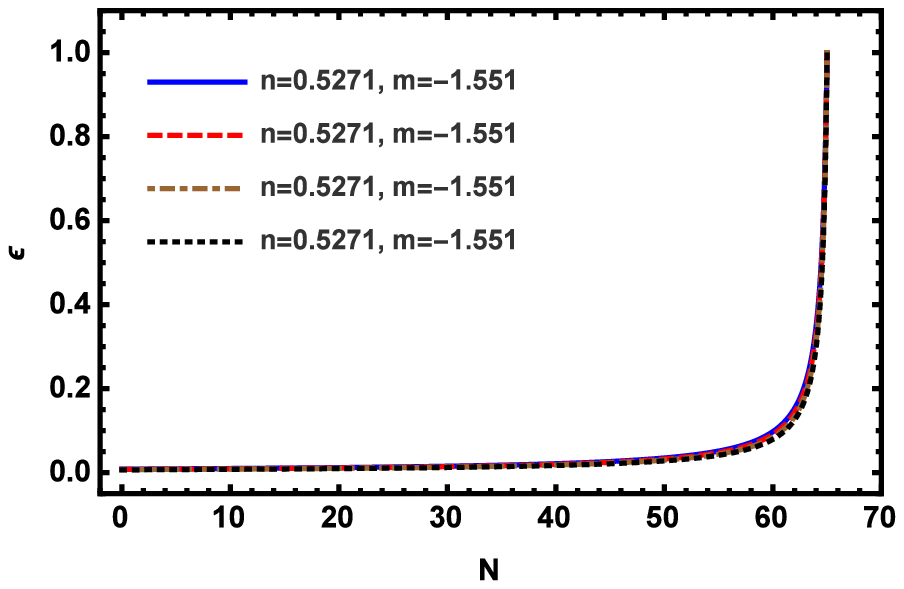}
\figcaption{\label{epsilonC01} Behavior of the slow-roll parameter $\epsilon$ versus the number of e-fold for different values of $n$ and $m$ selected from Fig.\ref{SnmQTH01}. }
\end{center}

\subsection{Second Case: $\Gamma := \Gamma(\phi,T)$}
In this section we consider a more generalize case so that the dissipation coefficient is a function of both the tachyon field and the temperature. A common choice is \cite{Zhang:2009ge,Bastero-Gil:2014oga,Panotopoulos:2015qwa}
\begin{equation}\label{Gamma02}
\Gamma=\Gamma_0\dfrac{T^m}{\phi^{m-1}}.
\end{equation}
The temperature of the fluid could be found in terms of the tachyon field from the following relation
\begin{equation}\label{T02}
  \rho_r = \alpha T^4 = {3 \over 4} \; \Gamma H \; \dot{\phi}^2,
\end{equation}
where the time derivative of the field is obtained from Eq.\eqref{phidot}. Using the definition of the dissipation coefficient, the temperature is expressed versus the field as
\begin{equation}\label{temp02}
  T^{m+4} = {3 n^2 \over \alpha} \; {H_0 \over \Gamma_0} \; \phi^{m+n-3}.
\end{equation}
Inserting this result in the definition of $\Gamma$, Eq.\eqref{Gamma02}, the parameter is read in terms of the scalar field as
\begin{equation}\label{Gammaphi}
  \Gamma = \bar{\Gamma}_0 \; \phi^b
\end{equation}
where
\begin{equation*}
  \bar{\Gamma}_0 \equiv \Gamma_0 \; \left( {3 n^2 \over \alpha} \; {H_0 \over \Gamma_0} \right)^{m \over m+4}, \qquad
  b \equiv {nm-6m+4 \over m+4},
\end{equation*}
The scalar field at the end of inflation is derived from the relation $\epsilon = 1$, which indicates the end of acceleration expansion phase. Then, following the same process as the previous case, the scalar field at the time of horizon crossing is achieved, so that
\begin{eqnarray}\label{phis02}
\phi_\ast^{b+n+2} &= & \phi_e^{b+n+2} \; \left( 1 + {b+n+2 \over n} \; N \right) \equiv \phi_e^{b+n+2} \tilde{N}, \quad \nonumber \\
\phi_e^{b+n+2} &= & {2n^2 \over \bar{\Gamma}_0 H_0}.
\end{eqnarray}
Substituting above relation in the definition of the slow-roll parameters, one could compute these parameters at the horizon crossing time in terms of the number of e-fold as
\begin{eqnarray}
  \epsilon^\ast &=& \tilde{N}^{-1} \\
  \eta^\ast &=& {n-1 \over n} \; \epsilon^\ast \\
  \beta^\ast &=& {b \over n} \; \epsilon^\ast
\end{eqnarray}
Inserting the above parameters in \eqref{nssdr}, the scalar spectral index is obtained
\begin{equation}\label{nssdr02}
  n_s - 1 = \left( {13 \over 4}  + {3 \over 2} \; {n-1 \over n} + {7 \over 4} \; {b \over n} \right) \; \tilde{N}^{-1},
\end{equation}
which states that the parameter depends on the constants $n$ and $m$. The power-spectrum of the scalar perturbation and tensor-to-ratio have the same form as they were obtained in Eqs.\eqref{pssdr01} and \eqref{rsdr01} where $\Gamma_0$ and $m$ are respectively replaced by $\tilde{\Gamma}_0$ and $b$. \\
Same as the first case, the energy scale of the inflation is utilized to determine the constant $H_0$, which comes to the following expression
\begin{equation}\label{H0-02}
  H_0 = \tilde{V}^{\star} \; \Gamma_0^{-n \over m-3}, \qquad
  \tilde{V}^{\star} \equiv { \left( V^\star / 3 \right)^{nm+2n-2m+6 \over -4(m-3)} \over \left[ 2n^2 \tilde{N} \left( {\alpha \over 3n^2} \right)^{m \over m+4} \right]^{-n(m+4) \over 4(m-3)} }
\end{equation}
Applying above relation in the power-spectrum of the scalar perturbation, and compute the power-spectrum for $t=t_\star$, one could specify the other constant of the model, so that
\begin{equation}\label{G0-02}
  \Gamma_0 = \left( \mathcal{P}_s^\star \over \tilde{D} \right)^{1/g}
\end{equation}
where the defined constants are expressed as
\begin{eqnarray*}
  \tilde{D} &\equiv& {\left( {3n^2 \over \alpha} \right)^{m \over 4(m+4)} \tilde{V}^{\star f} \over 8\pi^2 n^{3/2} (3\alpha)^{1/4}} \;
  {\exp\left[ {-3 (2n+b) \over 8(b+n+2) \; \tilde{N}} \right] \over \left( 2n^2 \tilde{N} \left( {\alpha \over 3n^2} \right)^{m \over m+4} \right)^{q}} \\
  q &\equiv& {7 b +19n-6 \over 4(b+n+2)} \\
  f &\equiv& {2q(m+2)-1 \over m+4} \\
  g &\equiv& {4q+1 \over m+4} - {n f \over m-3}
\end{eqnarray*}
Inserting Eqs.\eqref{H0-02} and \eqref{G0-02}\emph{ into} the tensor-to-scalar ratio gives
\begin{equation}\label{rssdr02}
  r^\star = { 4 n^{3/2} (3\alpha)^{1/4} \over \left( {3n^2 \over \alpha} \right)^{m \over 4(m+4)}} \;
           {\left( 2n^2 \tilde{N} \left( {\alpha \over 3n^2} \right)^{m \over m+4} \right)^{p} \over \exp\left[ {-3 (2n+b) \over 8(b+n+2) \; \tilde{N}} \right]} \;
           {H_0^{8m-2p(m+2)+36 \over 4(m+4)} \over \Gamma_0^{4p \over m+4}}
\end{equation}
in which the \emph{parameter} $p$ is defined as
\begin{equation*}
  p \equiv {7b+27n-6 \over 4(b+n+2)}.
\end{equation*}
From Eqs.\eqref{H0-02} and \eqref{G0-02}, it is clear that the tensor-to-scalar perturbation \eqref{rssdr02} depends on the constants $n$ and $m$. On the other hand, the scalar spectral index \eqref{nssdr02} only depends on these two constants. Using the Planck $r-n_s$ diagram, the valid values of $n$ and $m$ are clarified in which for them the model prediction about the scalar spectral index and tensor-to-scalar ratio perfectly meet the observational data. These values have been plotted in Fig.\ref{Snm02}.  \\
\begin{center}
\includegraphics[width=8cm]{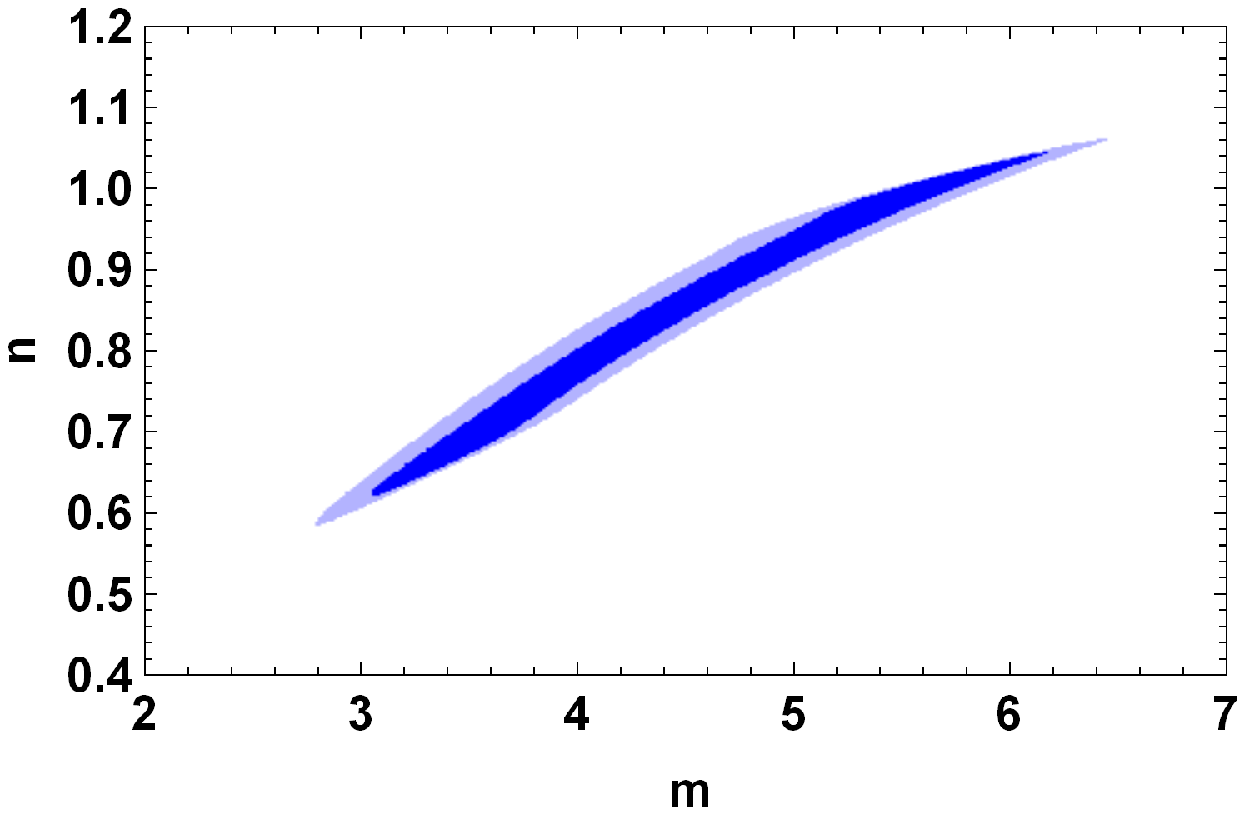}
\figcaption{\label{Snm02} The parametric plot of $(n,m)$.}
\end{center}
The next step is to examine whether this values of $(n,m)$ are consistent with the assumptions, i.e. $T/H>1$ and $Q\gg 1$, that were used for building the model. The fluid temperature is given in Eq.\eqref{temp02}, and the dissipative parameter for the case is read as
\begin{equation}\label{Q02}
  Q = {\Gamma_0 \over 3H_0} \; \left( {3n^2 \over \alpha} \; {H_0 \over \Gamma_0} \right)^{m \over m+4} \; {1 \over \phi^{2(2n+3m-2) \over m+4}}.
\end{equation}
Inserting $\phi_\star$ and using Eqs.\eqref{H0-02} and \eqref{G0-02}, both temperature and the dissipative parameter are expressed in terms of the constants $n$ and $m$. Further investigation \textbf{determines} that the obtained range of $(n,m)$, that has been plotted in Fig.\ref{Snm02}, could perfectly satisfy both conditions. To have a better insight, Figs.\ref{THQ02a} and \ref{THQ02b} respectively display the behavior of $T/H$ and $Q$ for different choices of $n$ and $m$. The figures clearly display that $T/H$ increases by passing time and approaching to the end of inflation, while the dissipative parameter $Q$ completely shows a different behavior so that it begins from high values and then reduces. However, the most important point is that during the inflation, they are always much bigger than one and the conditions $T/H > 1$ and $Q > 1$ are perfectly satisfied.   \\
\begin{center}
\includegraphics[width=8cm]{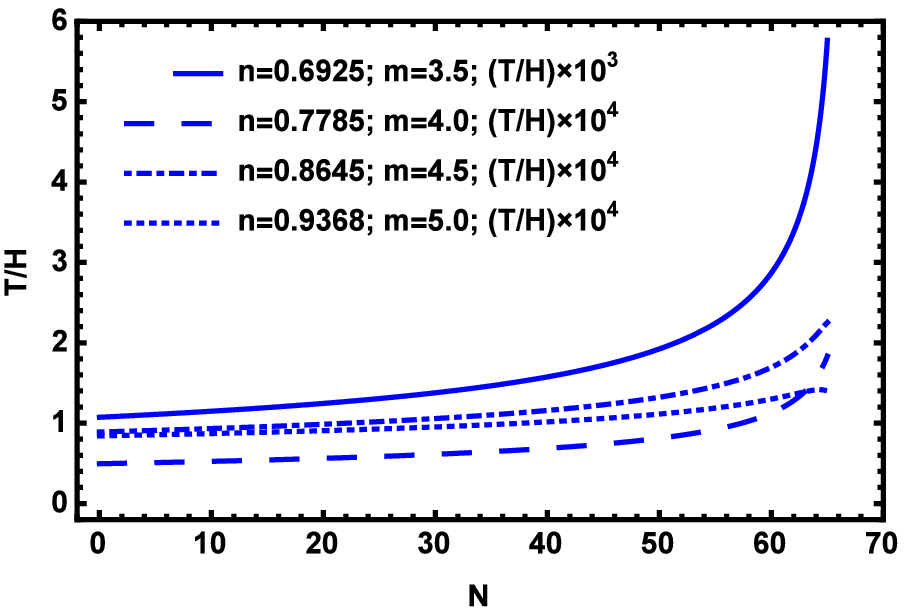}
\figcaption{\label{THQ02a} Behavior of $T/H$ during inflation for different values of $n$ and $m$ selected from Fig.\ref{Snm02}. The plots indicate that both conditions $T/H > 1$ and $Q \gg 1$ are perfectly satisfied.}
\end{center}
\begin{center}
\includegraphics[width=8cm]{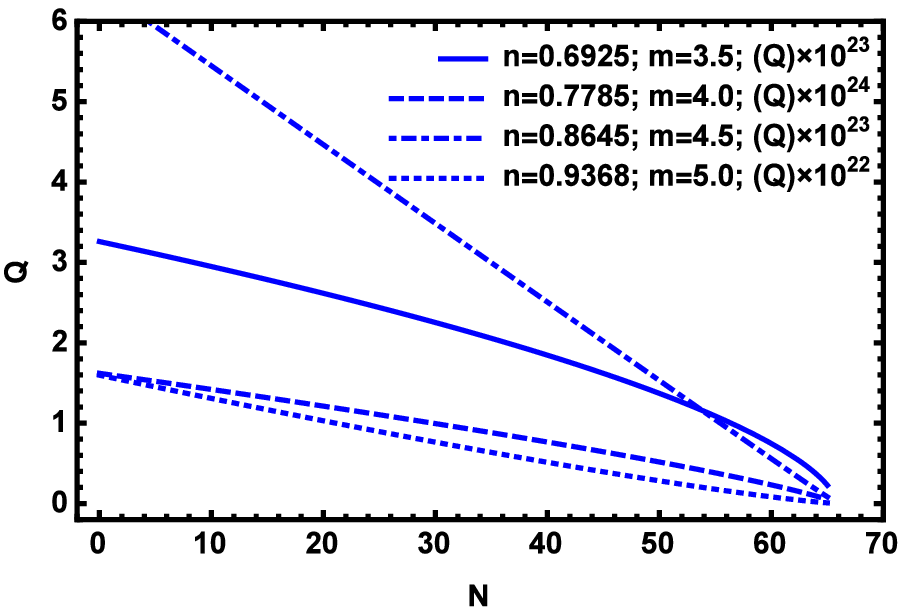}
\figcaption{\label{THQ02b} Behavior of $Q$ during inflation for different values of $n$ and $m$ selected from Fig.\ref{Snm02}. The plots indicate that both conditions $T/H > 1$ and $Q \gg 1$ are perfectly satisfied.}
\end{center}
To have a numerical insight about the result, the perturbation parameters of the model and also the temperature and dissipative parameter are presented in Table.\ref{Ts02}, where they are found out for different values of $n$ and $m$. \\
\end{multicols}

\begin{center}
\tabcaption{\label{Ts02} numerical results of the case.}
\footnotesize
\begin{tabular}{cccccccccc}
\hline
$n$ & $m$ & $\phi_\star$ & $\phi_e$ & $\Gamma_0$ & $H_0$ & $n_s$ & $r$ & $T/H$ & $Q$ \\
\hline
$0.6547$ & $3.23$ & $3.74 \times 10^{-6}$ & $1.72 \times 10^{-8}$ & $1.06 \times 10^{19}$ & $3.26 \times 10^{-3}$ & $0.9703$ & $0.0161$ & $319.35$ & $1.04 \times 10^{23}$ \\
    $0.7175$ & $3.66$ & $5.13 \times 10^{-6}$ & $1.47 \times 10^{-8}$ & $2.10 \times 10^{18}$ & $5.69 \times 10^{-3}$ & $0.9662$ & $0.0046$ & $2019.82$ & $6.50 \times 10^{23}$ \\
    $0.7889$ & $4.02$ & $2.06 \times 10^{-6}$ & $5.23 \times 10^{-9}$ & $2.12 \times 10^{17}$ & $0.0279$ & $0.9675$ & $0.0005$ & $3782.75$ & $1.27 \times 10^{24}$ \\
    $0.8374$ & $4.35$ & $1.55 \times 10^{-6}$ & $3.06 \times 10^{-9}$ & $3.02 \times 10^{16}$ & $0.0667$ & $0.9665$ & $0.0002$ & $9406.08$ & $1.39 \times 10^{24}$ \\
    $0.9202$ & $4.89$ & $4.86 \times 10^{-7}$ & $7.48 \times 10^{-10}$ & $2.20\times 10^{14}$ & $0.5883$ & $0.9678$ & $0.0003$ & $10646.8$ & $5.99 \times 10^{22}$ \\
    $0.9630$ & $5.25$ & $3.38 \times 10^{-7}$ & $3.99 \times 10^{-10}$ & $9.89 \times 10^{12}$ & $1.5566$ & $0.9673$ & $0.0011$ & $12190$ & $3.31 \times 10^{21}$ \\
    \hline
    $2$ & $1$ & $6.55$ & $1.93$ & $2.04 \times 10^{11}$ & $2.12 \times 10^{-8}$ & $1.0305$ & $6.00 \times 10^{-27}$ & $396.85$ & $7.26 \times 10^{9}$ \\

    $1$ & $-1$ & $3.98$ & $1.43$ & $3.71 \times 10^{2}$ & $2.34 \times 10^{-7}$ & $1.0217$ & $8.69 \times 10^{-15}$ & $658.79$ & $3.32 \times 10^{14}$ \\

    \hline
  \end{tabular}
\end{center}

\begin{multicols}{2}
%
The usual values of $n$ and $m$, that we have in the canonical cases, are listed in the last two rows of the table \ref{Ts02}. Based on Fig.\ref{Snm02}, these points are out of our interested range and the results are expected to be not in consistency with observation. Table.\ref{Ts02} represents the numerical results and states that although the predicted $r$ agrees with data, the result for the scalar spectral index is larger than unity, and clearly in tension with data. Then, these values of $n$ and $m$ are not suitable for the presented model.  \\

To check the graceful exit of inflation, the evolution of the first slow-roll parameter $\epsilon$ is investigated and plotted in Fig.\ref{epsilonC02}. It is realized that $\epsilon$ increases by approaching to the end of inflation, and eventually it reaches to one. Therefore, inflation ends at this time and the universe exits from the inflationary stage.
\begin{center}
\includegraphics[width=8cm]{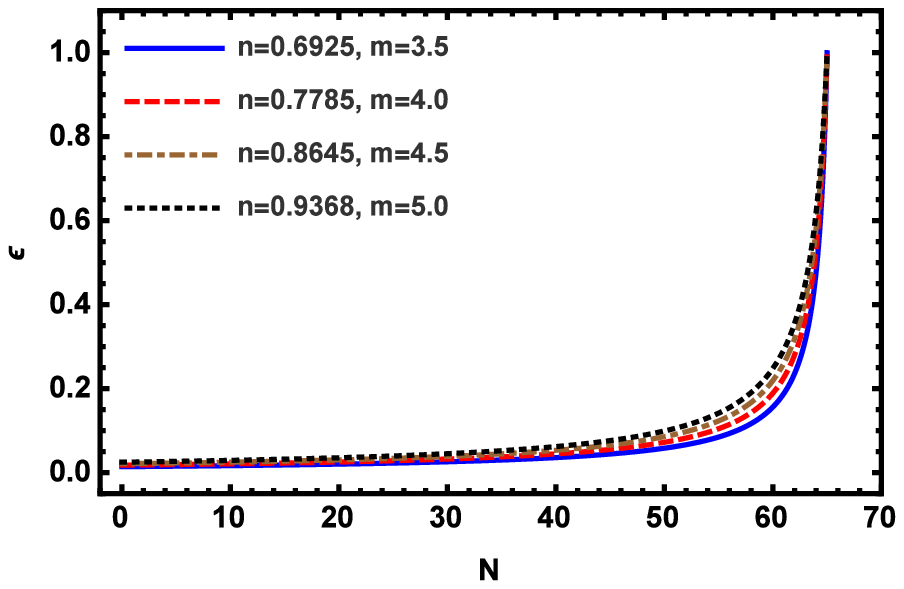}
\figcaption{\label{epsilonC02} Behavior of the slow-roll parameter $\epsilon$ versus the number of e-fold for different values of $n$ and $m$ selected from Fig.\ref{Snm02}. }
\end{center}

\section{Discussing the swampland criteria}\label{Swampland}
One of the best candidates of quantum gravity may is string theory which provides a landscape containing consistent low-energy EFTs that could formulate a quantum theory. However, \textbf{all other low-energy EFTs} live on a bigger region known as swampland. The EFTs which live on swampland are in contradiction with string theory. The desire for building a model based on the consistent EFT, which lives on the landscape, requires a mechanism to separate the consistent and inconsistent EFTs. The efforts have \textbf{resulted} in some conjectures in which the swampland criteria are the most recent proposal. The swampland criteria have been introduced in \cite{Obied:2018sgi}, and then it has been refined in \cite{Ooguri:2018wrx}. In brief, they are as follows:
\begin{itemize}
  \item {\bf C1: The distance conjecture:} it is an upper bound that confines the scalar field excursion in the field space as \begin{equation}\label{c1}
           \Delta \phi \leq \delta \sim \mathcal{O}(1).
         \end{equation}
  \item {\bf C2: The de Sitter conjecture:} it imposes a lower bound on the gradient of the potential stating that slope of a positive potential, $V>0$, of the scalar field should satisfy the following bound
      \begin{equation}\label{c2}
        {|V_\phi| \over V} \geq c \sim \mathcal{O}(1).
      \end{equation}
      and the refined version of this conjecture is given by
      \begin{equation}\label{c2refined}
        {|V_\phi| \over V} \geq c \sim \mathcal{O}(1), \quad {\rm or} \quad  {V_{\phi\phi} \over V} \geq -c' \sim -\mathcal{O}(1).
      \end{equation}
\end{itemize}
Note that we are working in Planck units where $M_p =1$. The exact value of the constant $c$ depends on the detail of the compactification which states that it could be larger than $\sqrt{2}$. However, further investigation shows that it could be smaller than unity, even of order $\mathcal{O}(0.1)$, and the important point is that it should be positive \cite{Kehagias:2018uem}. \\

It is believed that inflation occurs at the energy level below than the Planck energy scales, where it could properly be described by low-energy EFT \cite{Das:2018hqy,Das:2018rpg,Goswami:2019ehb}. Therefore, it is our interest to build the inflationary model in the framework of a consistent low-energy EFT that stands in landscape. In this regard, despite having \textbf{an agreement} with the observational data, which have been investigated previously, the inflationary model should also satisfy two swampland criteria. In the previous sections, the warm inflationary scenario was considered in SDR where the tachyon field had the role of \textbf{the inflaton}. In the previous sections, in comparison with the observational data, the constants of the model were determined. Now, we are about to find out whether the obtained results put the model in consistency with the swampland criteria. \\
In the first case, where the dissipation coefficient is picked out as a power-law function of the scalar field, \textbf{a narrow range} is obtained for the constants $n$ and $m$ in which only for these values the model comes to an agreement with observational data. However, for these values of $n$ and $m$, the difference of the scalar field at the time of horizon crossing and end of inflation is of order $\mathcal{Q}(10)$ or sometimes even larger, i.e. $\mathcal{Q}(10^2)$; it is clear from the Table.\ref{Ts01}. Therefore, it could be concluded that although the model is in good consistency with observational data, it does not satisfy the first swampland criterion. The result is different for the second case of 
\textbf{the dissipation } coefficient, where the parameter $\Gamma$ is a function of both the scalar field and temperature. The determined values of $n$ and $m$ state that the scalar field values at the end of inflation and also at the beginning of the horizon crossing time are smaller than unity, in which Table.\ref{Ts02} shows this conclusion, which in turn indicates that the field excursion during inflation is smaller than unity. Therefore, the first swampland criterion is satisfied for the second case of the presented model. \\

The second criterion has received more attention in the \textbf{works of literature} since it seems to be in direct tension with one of the fundamental assumptions of the standard inflationary scenario. The standard inflationary scenario is usually explained by means of the slow-roll parameters. The slow-roll parameter $\epsilon$ is defined as $\epsilon \simeq V'^2 / 2 V^2$, which should be smaller than unity to have accelerated expansion phase. On the other hand, based on the second swampland criterion it should be larger than a constant $c$ that is of order of unity, however, $c \approx 0.1$ could also work properly. Taking the latent value, the slow-roll parameter $\epsilon$ is obtained as $\epsilon = 0.005$ (for the best case) which is small enough to give a desire accelerated expansion phase. But the problem encounters when we are going to examine the tensor-to-scalar ratio $r$ with data. Based on the standard inflation model, the parameter is given by $r=16 \epsilon$ which for the aforementioned value of $\epsilon$ it is acquired about $r=0.08$ that is in tension with observational data. The problem might be \textbf{solved} for the generalized model of inflation such as k-essence and multi-field inflation. In the k-essence model of inflation the tensor-to-scalar ratio is modified as $r=16 c_s \epsilon$ where $c_s$ is the sound speed that could be of order of $0.1$ \cite{Kehagias:2018uem}. In addition, warm inflation, especially when the model is considered in SDR, could suits the swampland criteria. In the warm inflation, the first-slow-roll parameter is generalized as $\epsilon = \epsilon_\phi / Q$ where $\epsilon_\phi$ is the same slow-roll parameter that we have in cold inflation, i.e. $\epsilon_\phi = V'^2 / 2 V^2$ and $Q$ is the dissipative parameter which is much larger than unity in SDR. Then, the second swampland \textbf{criterion} implies that there should be $\epsilon_\phi = Q \epsilon > c^2/2$. Here, we worked with the tachyon field as the inflaton, where the first slow-roll inflation is given by Eq.\eqref{epsilon} in terms of the Hubble parameter. The potential of the field is related to the Hubble parameter through the relation $V(\phi) = 3H^2 \sqrt{1 - 2 \epsilon /3Q}$. Since, the first slow-roll parameter $\epsilon$ is small also dues to the fact that we are working in SDR, the last term could be ignored in which with a good approximation we have $V(\phi) = 3 H^2$. Therefore, the gradient of the potential is given by $V' / V = 2 H' / H$, and by applying Eq.\eqref{epsilon} the second criterion is read as
\begin{equation}\label{c2tachyon}
  {V' \over V} \simeq 2 {H' \over H} = \left( 6 Q H^2 \epsilon \right)^{1/2}.
\end{equation}
According to the second swampland criterion, \textbf{the gradient} of the potential should be larger than the constant $c$ that is of order of unity. From Table.\ref{Ts01}, which determines the values of the parameters of the model for the first case of the dissipation parameter, it seems unlikely to arrive at consistency between the model and the criterion. To have a better understanding, Fig.\ref{c201} displays the gradient of the potential versus the number of e-fold $N$ from the beginning of inflation to the end. It clearly indicates that the potential gradient increases by approaching to the end of inflation however it never reaches to one. \\
\begin{center}
\includegraphics[width=8cm]{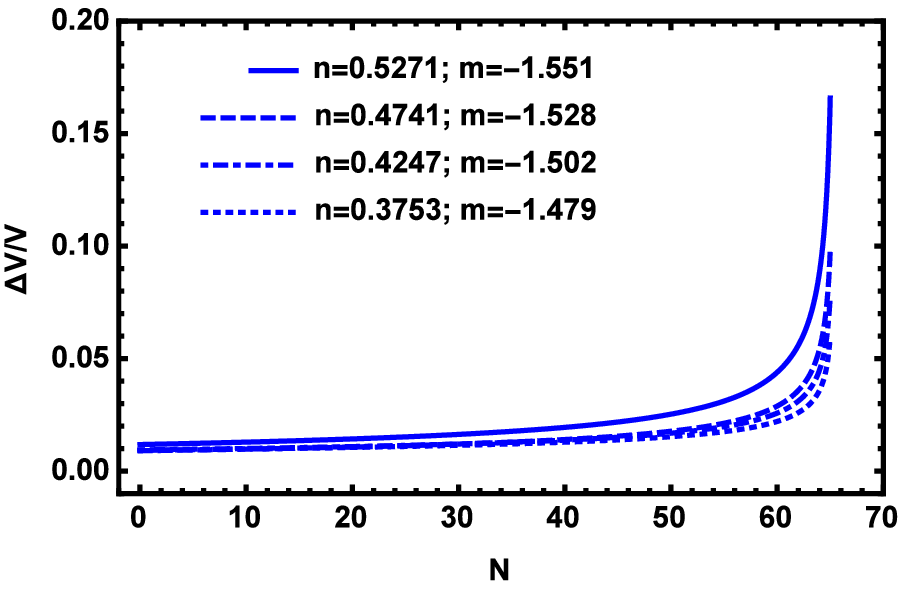}
\figcaption{\label{c201} The behavior of $\Delta V/V$ versus the number of e-fold during the inflation.}
\end{center}
The situation is different for the second case that is mostly because of the high value of the dissipative parameter $Q$ for the case. The potential gradient of the field for this case is depicted in Fig.\ref{c202}, where it is clearly realized that $\Delta V/V$ is clearly bigger than one during the whole time of inflation and the second swampland criterion could be properly satisfied. \\

\begin{center}
\includegraphics[width=8cm]{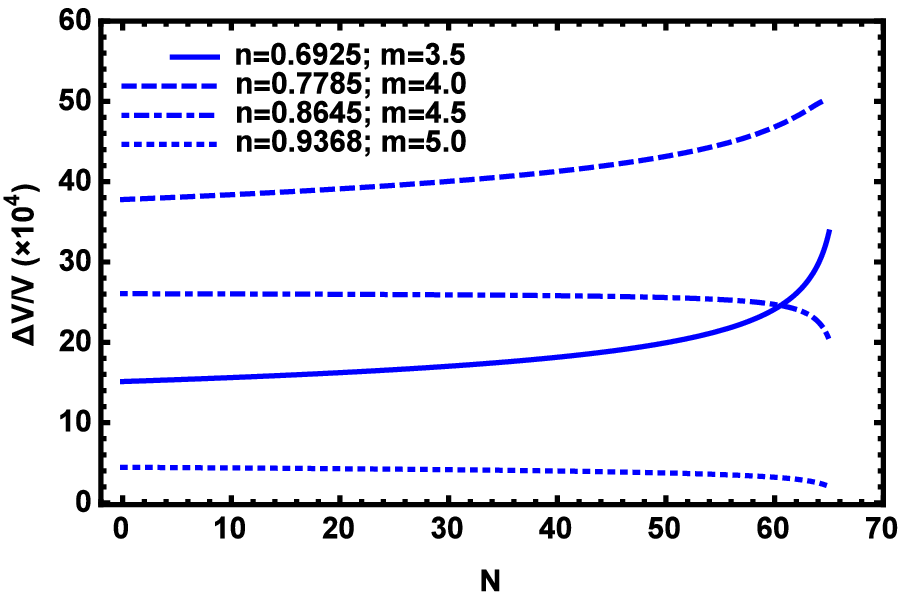}
\figcaption{\label{c202} The behavior of $\Delta V/V$ versus the number of e-fold during the inflation.}
\end{center}

In Brief, the first choice of the dissipation coefficient for the tachyon scalar field could be in great agreement with observational data, however, it could satisfy none of the swampland criteria. On the other hand, the second choice of the dissipation coefficient provides our desire result. It could come to an agreement with observational data and at the same time it could properly satisfy the swampland criteria.

\section{Conclusion}
The scenario of two components warm inflation including the tachyon field as the inflaton and photon gas was considered. The \textbf{Universe} is assumed to be filled with the scalar field and radiation in which they interact with each other and energy is transferred from the scalar field to radiation. The interaction term includes a dissipation coefficient besides the sum of the energy density and the pressure of the inflaton, which for the case of the standard model of scalar field it \textbf{goes} back to the familiar term $\Gamma \dot{\phi}^2$. This type of interaction will be different for each model, however it comes to the same dissipative parameter, $Q=\Gamma / 3H$ regardless of the type of the scalar field model. \\
Warm \textbf{inflation} scenario usually is considered in two different regimes as \textbf{weak and strong} dissipative regimes,  respectively correspond to $Q \ll 1$ and $Q \gg 1$. The work was restricted only to the strong dissipative regime. Imposing this assumption, the main dynamical and perturbation parameters for the model were derived. To examine the validity of the model two different choices of the dissipation coefficient were studied. The dissipation coefficient is a function of the scalar field or temperature or in some cases both of them. In this present work, two different choices were taken \textbf{into} account for $\Gamma$. As the first case, it was as a function of the tachyon field, and in the second case, a more general case was considered, i.e. a function of both tachyon field and temperature. \\
The model was investigated in \textbf{detail} for both cases and the free parameters of the model were determined using the observational data. By calculating the slow-roll parameters at the horizon crossing, the scalar spectral index was obtained in terms of the constants $n$ and $m$. Then, using the energy scale of inflation and the amplitude of the scalar perturbations, the other constants $H_0$ and $\Gamma_0$ were determined. Using these results, It was found that at the horizon crossing the tensor-to-scalar ratio only depends on $n$ and $m$, too. Comparing the theoretical results with the Planck $r-n_s$ diagram, we found a set of $(n,m)$ points in which for them the model could perfectly meet the data. But, to get the ultimate consistent results, the validity of the first assumptions of the model should also be investigated. It was assumed that the thermal fluctuation dominates the quantum fluctuations, i.e. $T/H > 1$, and it was also supposed that inflation occurs in SDR, i.e. $Q>1$. Therefore, besides considering the consistency of the model with data, the self-consistency of the model was also considered. We tried to realize if the obtained set of $(n,m)$ points could guarantee the conditions. Examining these conditions for the first case demonstrated that the conditions are violated for some of the point of the set. There are only a few points that could guarantee the assumptions of the model and simultaneously put the model in agreement with data. The situation, however, is better for the second case, in which for the whole obtained $(n,m)$ points the conditions are fulfilled and also the results about the scalar spectral index and tensor-to-scalar ratio are in good agreement with data.  \\
The final part of the work was devoted to the recently proposed swampland conjectures. It is believed that any inflationary model should be in consistency with them, although they are not completely approved yet. There are two conditions that put an upper bound on the distance of the scalar field and the second condition imposes a lower bound on the gradient of the potential of the scalar field. The criteria could rule out some of the inflationary models, however, there is a strong belief that the warm inflation is able to properly satisfy the criteria; mostly because of the presence of the dissipative parameter $Q$ that is large in SDR. However, to have a precise conclusion, the model should be examined quantitatively. In this regard, both cases of the dissipation coefficient were examined which determined that the first case could not satisfy even one of the criteria. On the other hand for the second case, where the dissipation coefficient is a function of both the scalar field and the temperature, the model properly satisfies both swampland criteria. \\

\acknowledgments{
The work of A.M. has been supported financially by 'Vice Chancellorship of Research and Technology, University of Kurdistan'under research Project No.98/10/34704. The work of T. G. has been supported financially by 'Vice Chancellorship of Research and Technology, University of Kurdistan'under research Project No.98/11/2724. HS thanks A. Starobinsky for very constructive discussions about inflation during Helmholtz International Summer School  2019 in Russia. He grateful G. Ellis, A. Weltman, and UCT for arranging his short visit, and for enlightening discussions about cosmological fluctuations and perturbations for both large and local scales. He also thanks  T. Harko and H. Firouzjahi for constructive discussions about inflation and perturbations. His special thanks go to his wife E. Avirdi for her patience during our stay in South Africa.  }

\end{multicols}

\vspace{-1mm}
\centerline{\rule{80mm}{0.1pt}}
\vspace{2mm}

\begin{multicols}{2}


\bibliography{WTRef}


\end{multicols}

\clearpage
\end{CJK*}
\end{document}